\documentclass[twocolumn,secnumarabic,amssymb, nobibnotes, superscriptaddress, nofootinbib,aps,prd]{revtex4-2}
\usepackage[colorlinks,citecolor=blue,linkcolor=blue,anchorcolor=blue,filecolor=blue, urlcolor=blue]{hyperref}
\usepackage{natbib}

\usepackage{amsmath}
\usepackage{adjustbox}
\usepackage{graphicx}
\usepackage{booktabs}       
\usepackage{amsfonts}      
\usepackage{nicefrac}       
\usepackage{microtype}
\usepackage{cleveref}
\usepackage{xcolor}
\usepackage{multirow}
\usepackage{makecell}
\usepackage{array}
\usepackage{orcidlink}
\usepackage{float}
\usepackage{lineno}

\begin{document}

\author{Adamu Issifu \orcidlink{0000-0002-2843-835X}} 
\email{ai@academico.ufpb.br}
\affiliation{Departamento de F\'isica, Instituto Tecnol\'ogico de Aeron\'autica, DCTA, 12228-900, S\~ao Jos\'e dos Campos, SP, Brazil} 
\affiliation{Laborat\'orio de Computa\c c\~ao Cient\'ifica Avan\c cada e Modelamento (Lab-CCAM),  SP, Brazil}
\affiliation{CFisUC, Department of Physics, University of Coimbra, 3004-516 Coimbra, Portugal}

\author{Andreas Konstantinou~\orcidlink{0000-0002-1072-7313}}\email{akonst29@ucy.ac.cy}
\affiliation{Department of Physics, University of Cyprus, P.O. Box 20537, 1678 Nicosia, Cyprus}
\affiliation{Computation-based Science and Technology Research Center,
The Cyprus Institute, 20 Kavafi Str., Nicosia 2121, Cyprus}

\author{Franciele M. da Silva \orcidlink{0000-0003-2568-2901}} 
\email{franciele.m.s@ufsc.br}
\affiliation{Departamento de F\'isica, CFM - Universidade Federal de Santa Catarina, \\ Caixa Postal 5064, CEP 880.35-972, Florian\'opolis, SC, Brazil.}
\affiliation{Theoretical Astrophysics, Institute for Astronomy and Astrophysics, University of T\"{u}bingen, 72076 T\"{u}bingen, Germany}

\author{Tobias Frederico \orcidlink{0000-0002-5497-5490}} 
\email{tobias@ita.br}
\affiliation{Departamento de F\'isica, Instituto Tecnol\'ogico de Aeron\'autica, DCTA, 12228-900, S\~ao Jos\'e dos Campos, SP, Brazil}
\affiliation{Laborat\'orio de Computa\c c\~ao Cient\'ifica Avan\c cada e Modelamento (Lab-CCAM), SP, Brazil}

\title{Rotational effects in QSs: comparing different models}

\begin{abstract}
 We investigate the rotational properties of strange quark stars using two representative quark matter (QM) equations of state (EOS): the vector MIT bag model and the density-dependent quark mass (DDQM) model. Through general-relativistic calculations of uniformly rotating sequences, we analyze their mass--radius relations, moments of inertia, quadrupole moments, surface redshifts, Keplerian frequencies, and energy components. A central result of this work is the full decomposition of the stellar energy budget in rotating strange stars, separating gravitational, internal, rotational, and binding energy contributions. Rotation amplifies the intrinsic EOS differences: the MIT model supports more massive ($M_{\max} \gtrsim 3.3\,M_\odot$) compact stars with larger moments of inertia and greater resistance to deformation, while the DDQM model produces larger radii, less massive stars limited by low Keplerian frequencies. Combined measurements of mass, radius, moment of inertia, surface redshift, and spin frequency can thus break the EOS degeneracy; massive, rapidly rotating pulsars favor MIT-like EOS, whereas larger moment of inertia in canonical stars point to a DDQM-like model. Moreover, by contrasting the rotational properties of quark stars (QSs) with those of hadronic neutron stars (NSs), we identify distinctive multimessenger signatures of strange stars. These rotational observables, soon to be tightly constrained by Neutron Star Interior Composition Explorer (NICER) and next-generation gravitational-wave detectors, offer a means to test the existence and composition of QM in compact stars.
\end{abstract}

\maketitle

\section{Introduction}

The theoretical proposition that strange QM may constitute the true ground state of nuclear matter~\cite{Witten:1984rs} provides a compelling motivation for studying strange stars \cite{1986ApJ...310..261A}, compact objects composed of up, down, and strange quarks \cite{Issifu_20251, Issifu:2023ovi, daSilva:2023okq}. A definitive identification of such objects requires confronting theoretical predictions with astrophysical observations, many of which probe the properties of rapidly rotating compact stars. Early investigations of rotating strange stars primarily employed the vector MIT bag model EOS~\cite{LattimerApJ1990, StergioulasA&A1999, ZdunikA&A2000}, while subsequent studies introduced alternative descriptions, including density-dependent EOSs that serve as precursors to the more sophisticated models used here~\cite{BombaciApJL2000, Gondek-RosinskaA&A2000}.

Initial models of rotating strange stars, often based on simplified EOSs neglecting quark masses and interactions, predicted Keplerian frequencies comparable to or even lower than those of NSs~\cite{LattimerApJ1990, ZdunikPhRvD1990, PrakashPhLB1990}, a consequence of their relatively low maximum masses. However, modern EOSs that incorporate quark masses and interactions, including those employed in this study, yield substantially higher maximum masses and rotational frequencies~\cite{Gondek-RosinskaA&A2000, BhattacharyyaMNRAS2016}. The rotational properties of these QM configurations differ markedly from their NS counterparts. For instance, uniform rotation can enhance the maximum mass of a strange star by over $40\%$~\cite{GourgoulhonA&A1999, Gondek-RosinskaA&A2000, ZhouAN2017}, compared with a typical $\sim 20\%$ increase for NSs, while differential rotation can support even more massive configurations~\cite{SzkudlarekASPC2012, SzkudlarekApJ2019, ZhouPhRvD2019}. Furthermore, the ratio of rotational kinetic to gravitational binding energy, $T/|W|$, can be significantly larger for strange stars~\cite{GourgoulhonA&A1999, Gondek-RosinskaA&A2000}, indicating an increased susceptibility to non-axisymmetric instabilities, though triaxial evolution may ultimately limit their attainable spin rates~\cite{ZhouJPhCS2017, ZhouPhRvD2018}.

A variety of observational signatures may distinguish strange stars from NSs, including quasi-periodic oscillations associated with orbital and epicyclic frequencies~\cite{Gondek-RosinskaA&A2001, Gondek-RosinskaAIPC2005, Gondek-RosinskaPhRvD2014}, properties of the innermost stable circular orbit \cite{StergioulasA&A1999, ZdunikA&A2001, BhattacharyyaA&A2001, Gondek-RosinskaA&A2001}, and universal relations linking the mass, radius, moment of inertia, and quadrupole moment {for static configurations~\cite{ChengPhRvD2000, DonevaPhRvD2014, ChenPhRvD2023}. The strong bulk viscosity of strange quark QM suppresses the r-mode instability, suggesting that the detection of a sub-millisecond pulsar would constitute compelling evidence for strange stars~\cite{MadsenPhRvL1998}, a possibility further explored in subsequent r-mode analyses~\cite{Madsen2000PhRvL, Andersson2002MNRAS, Gondek-RosinskaA&A2003}. 
Although a larger $T/|W|$ ratio could imply enhanced gravitational-wave emission, simulations indicate that triaxially rotating QSs may not be substantially more efficient continuous-wave sources than NSs~\cite{ZhouPhRvD2018, ZhouPhRvD2021}.

 In this work, we investigate the rotational effects on the structural properties of bare, self-bound strange stars composed of $u$, $d$, and $s$ quark species without a crust~\cite{Melrose:2006zg, Kettner:1994zs}. We perform the first fully relativistic comparison of two modern and widely used QM EOS whose free parameters are constrained using the same set of astrophysical data: the DDQM model and the vector MIT bag model. Using general-relativistic sequences of uniformly rotating configurations, we compute and contrast key stellar observables, including the mass–radius relation, moment of inertia, quadrupole moment, surface redshift, and, importantly, a complete decomposition of the stellar energy budget, providing complementary knowledge to advance observational missions.  We present the first full energy decomposition for rotating QSs, providing new and independent constraints on EOS stiffness and self-binding. 

Our goal is to determine how these rotational observables, now increasingly accessible through NICER and future gravitational-wave measurements, can be used to distinguish between competing QM descriptions. This analysis reveals clear EOS-dependent trends that link microphysical interactions to the macroscopic structure and stability of rotating strange stars. For rotating strange stars with a crust, we refer the reader to~\cite{GlendenningApJ1992, ZdunikA&A2001, HaenselA&A2009}. In addition to comparing the two QM models, we also contrast our results with the widely studied hadronic EOSs. This comparison enables the identification of distinctive multimessenger signatures that may help differentiate strange stars from hadronic NSs.

The paper is organized as follows: in \Cref{fm}, we present the theoretical framework underlying this study. This section is divided into three subsections: in \Cref{ddqm}, we introduce the DDQM model; in \Cref{sec_MIT}, we describe the vector MIT bag model; and in \Cref{rns}, we outline the theory governing the rotational properties of strange stars. Our results and discussions are presented in \Cref{rd}, and we conclude with final remarks in \Cref{remarks}.

\section{Model Formalism}\label{fm}
This section is divided into three subsections to discuss the microscopic and macroscopic physics underlying the study comprehensively. The EOS parameterizations used here come from Ref.~\cite{daSilva:2023okq}, which employs Bayesian inference to optimize the free parameters of the DDQM and the vector MIT bag models. The DDQM and vector MIT bag models were chosen because they provide two complementary and physically distinct descriptions of QM. The DDQM model, with density-dependent quark masses, yields a relatively soft EOS and stars with larger radii and lower maximum masses. In contrast, the vector MIT bag model, incorporating repulsive vector interactions, produces a much stiffer, self-bound EOS capable of supporting compact and massive configurations. Together, these models span the plausible extremes of QM stiffness, from a relatively soft, gravitationally-bound state to a stiff, self-bound one. This allows us to systematically assess how microphysical differences translate into observable rotational and structural properties, and to interpret multimessenger constraints in the context of the strange matter hypothesis.

\subsection{Density-Dependent Quark Mass Model} \label{ddqm}

The DDQM model describes strange QM by incorporating quark interactions through a density-dependent quark mass \cite{Xia:2014zaa, daSilva:2023okq, Issifu_2025}:
\begin{equation}
    m_i(n_B) = m_{0i} + \frac{D}{n_B^{1/3}} + C n_B^{1/3},
\end{equation}
where \(m_{0i}\) denotes the current quark mass {(with the subscript \(i = u, d, s\) corresponding to the up, down, and strange quarks, respectively)}, and \(C\) and \(D\) are model parameters. {The current quark masses \(m_{0u} = 2.16\,\mathrm{MeV}\), \(m_{0d} = 4.67\,\mathrm{MeV}\), and \(m_{0s} = 93.4\,\mathrm{MeV}\) are taken from the PDG report \cite{ParticleDataGroup:2022pth}.} To ensure thermodynamic consistency when $m_i(n_B)$ depends on density, we introduce the effective chemical potentials $\mu_i^*$, such that the free-energy density takes the form \cite{Backes:2020fyw, Xia:2014zaa}:
\begin{equation}
    f = \Omega_0(\{\mu_i^*\},\{m_i(n_B)\}) + \sum_i \mu_i^* n_i,
\end{equation}
with the quark thermodynamic potential
\begin{equation}
    \Omega_0 = - \sum_i \frac{\gamma_i}{24\pi^2} \left[ \mu_i^* p_{Fi}\left(p_{Fi}^2 - \tfrac{3}{2}m_i^2\right) + \tfrac{3}{2}m_i^4 \ln\!\left(\frac{\mu_i^*+p_{Fi}}{m_i}\right) \right],
\end{equation}
where $\gamma_i=6$ (3 colors $\times$ 2 spins) because the model is flavor independent, and 
\begin{equation}
    p_{Fi} = \sqrt{\mu_i^{*2} - m_i^2}
\end{equation}
is the Fermi momentum. The number density is then
\begin{equation}
    n_i = \frac{\gamma_i p_{Fi}^3}{6\pi^2}.
\end{equation}
Since QSs are stable macroscopic systems, their matter must satisfy charge neutrality and chemical stability, which requires the inclusion of leptons treated as free Fermi gases. Hence, we introduce the electron \(e\), with a spin degeneracy of \(\gamma_e = 2\) and mass $m_e = 0.511\,\rm MeV$. The EOS is obtained by imposing $\beta$-equilibrium,
\begin{equation}
    \mu_u^* + \mu_e = \mu_d^* = \mu_s^*,
\end{equation}
charge neutrality,
\begin{equation}
    \frac{2}{3}n_u - \frac{1}{3}n_d - \frac{1}{3}n_s - n_e = 0,
\end{equation}
and baryon number density,
\begin{equation}
    n_B = \frac{n_u+n_d+n_s}{3}.
\end{equation}

The energy density and pressure are given by
\begin{align}
    \varepsilon &= \Omega_0 - \sum_i \mu_i^* \frac{\partial \Omega_0}{\partial \mu_i^*}, \\
    P &= -\Omega_0 + \sum_{i,j} \frac{\partial \Omega_0}{\partial m_j} \, n_i \frac{\partial m_j}{\partial n_i}.
\end{align}

\subsection{Vector MIT Bag Model} \label{sec_MIT}

The vector MIT bag model extends the original MIT framework~\cite{PhysRevD.9.3471} by incorporating repulsive vector interactions in analogy with Quantum Hadrodynamics~\cite{Serot:1997xg}, %{\color{red}TF: is that Eq.14, if so should be the self-interaction of the vector-field? 
including a Dirac-sea self-interaction term~\cite{Furnstahl:1996zm}. The full Lagrangian is~\cite{Lopes:2020btp, Lopes:2020dvs}:
\begin{equation}
\mathcal{L} = \mathcal{L}_{\mathrm{MIT}} + \mathcal{L}_{\mathrm{V}} + \mathcal{L}_{\mathrm{Dirac}}
\end{equation}
with
\begin{equation}
    \mathcal{L}_{\mathrm{MIT}} = \sum_{i}\left\{\bar{\psi}_i \left(i\gamma^\mu\partial_\mu - m_{0i}\right)\psi_i - B\right\}\Theta(\bar{\psi}_i\psi_i),
\end{equation}
\begin{equation}
    \mathcal{L}_{\mathrm{V}} = \sum_{i}\left\{\bar{\psi}_i g_{iV}\gamma^\mu V_\mu \psi_i - \frac{1}{2}m_V^2 V^\mu V_\mu\right\}\Theta(\bar{\psi}_i\psi_i),
\end{equation}
\begin{equation}
    \mathcal{L}_{\mathrm{Dirac}} = b_4\,\frac{(g^2 V_\mu V^\mu)^2}{4}.
\end{equation}
Here, $\psi_i$ is the quark field, $B$ is the bag pressure, $m_V$ is the vector meson mass, $g_{iV}$ is the quark–vector coupling, and $g \equiv g_{uV}$. The step function $\Theta(\bar{\psi}_i\psi_i)$ enforces confinement inside the bag.

Under the mean-field approximation~\cite{Menezes:2021jmw, Serot:1997xg}, the quark single-particle energy and vector mean field satisfy
\begin{align}
    E_i &= \sqrt{p_{Fi}^2 + m_i^2} + g_{iV}V_0, \\
    gV_0 + \left(\frac{g}{m_V}\right)^2 b_4\,(gV_0)^3 &= \left(\frac{g}{m_V}\right)^2 \sum_i g_{iV} n_i,
\end{align}
where $p_{Fi}$ is the Fermi momentum and $n_i$ is the quark number density.

The energy density and pressure are
\begin{align}
    \varepsilon_i &= \frac{\gamma_i}{2\pi^2}\int_0^{p_{Fi}} E_i \, p^2\, dp,\\
    \varepsilon &= \sum_i \varepsilon_i + B - \frac{1}{2}m_V^2 V_0^2 - b_4\,\frac{(g^2 V_0^2)^2}{4},\\
    P &= \sum_i \mu_i n_i - \varepsilon.
\end{align}
Here, $\gamma_i = 6$ ($3$ color states $\times$ $2$ spin states) for each quark flavor, reflecting the color-symmetric nature of the model. 
$G_V \equiv (g/m_V)^2$, and coupling ratio $g_{sV}/g_{uV}=0.4$~\cite{Lopes:2020btp}, note that the model assumes $g_{dV}=g_{uV}$. 

\subsection{Rotating NSs}\label{rns}
In this work, we use the 
\texttt{rns} code \cite{Stergioulas1995ApJ} to model the rigidly rotating strange stars.  
The code solves Einstein's field equations for a stationary and axisymmetric space-time
\begin{equation}
\begin{aligned}
  ds^2 = -e^{\tilde{\gamma} + \tilde{\rho}} dt^2 + e^{2\tilde{\alpha}} (dr^2 + r^2 d\theta^2) \\
  + e^{\tilde{\gamma} - \tilde{\rho}} r^2 \sin^2\theta (d\phi - \tilde{\omega} dt)^2,
\end{aligned}
\label{eq:metric}
\end{equation}
where \( \tilde{\gamma}, \tilde{\rho}, \tilde{\alpha}, \tilde{\omega} \) are metric potentials that depend on the radial coordinate \( r \) and the polar angle \( \theta \). 

The total gravitational mass, $M$, the baryon mass, $M_0$, the angular momentum, J, the circumferential radius at the equator, $R_e$, and the polar redshift, $Z_p$ are given by equations (B1), (B2), (B4), (B6), (B7) of \cite{Cook1994ApJ424}, respectively.

The moment of inertia will be given by I = J/$\Omega$, the kinetic energy is defined as $T={J\Omega}/{2}$, and the binding energy is obtained by $E_{\mathrm{bind}}=M-M_0$. Here, $\Omega=2\pi\nu$ with $\nu$ being the spin frequency in Hz.

The absolute value of the gravitational binding energy is given by {$|W| = |M - M_p - T|$}, where $M_p$ is the proper mass and is defined in equation (65) of \cite{komatsu1989MNRASa}, and the internal energy is given by $U=M_p-M_0$.

The mass quadrupole moment is evaluated following the asymptotic expansion of the metric functions at large distances.  
In this approach, the \texttt{rns} code extracts the coefficient $M_2$ from the $r^{-3} P_2(\cos\theta)$ term in the expansion of the $\tilde{\rho}$-potential \cite{Laarakkers_1999,pappas2012multipolemomentsnumericalspacetimes}.  
The physical quadrupole moment $Q$ is then obtained through
\begin{equation}
Q = M_2 - \frac{4}{3}\left(\frac{1}{4} + b\right) M^3,
\label{eq:Q-final}
\end{equation}
where $b$ is a dimensionless coefficient determined from the asymptotic behavior of the metric potentials (see Eq.~(9) in \cite{Laarakkers_1999}), and $M_2$ corresponds to the coefficient of the $l=2$ (quadrupole) term in the expansion (see Section 2 of \cite{pappas2012multipolemomentsnumericalspacetimes}).

The baryonic mass is given by
\begin{equation}
M_0 = 2 \pi \int e^{2\tilde{\alpha} + (\tilde{\gamma} - \tilde{\rho})/2}
\frac{m_N n_B}{\sqrt{1 - \upsilon^2}}
\, r^2 \sin\theta \, dr\, d\theta\, ,
\end{equation}
with \(m_N =  939 \,\rm MeV\) the nucleon mass, \(n_B\) the baryon density, and
\(\upsilon = (\Omega - \tilde{\omega}) r \sin\theta \, e^{-\tilde{\rho}}\).
The enhancement of the maximum gravitational mass due to rotation is defined as
\begin{equation}
\Delta M_{\max}(r_p/r_e) =
M_{\max}(r_p/r_e)  - M_{\max}(1.0),
\end{equation}
where $M_{\max}(1.0)$ denotes the maximum mass of the corresponding non-rotating configuration. The corresponding percentage increase can then be obtained as $\Delta M_{\max}/M_{\max}(1.0) \times 100$.

\section{Results}\label{rd}
\begin{table*}[t]
\caption{Astrophysical sources used to constrain the DDQM and vector MIT bag models. We list the source type, spin frequency $\nu$, and the inferred mass and radius where available.}
\begin{ruledtabular}
\begin{tabular}{l c c c c}
\textbf{Star} & \textbf{Type} & $\boldsymbol{\nu}$ \textbf{(Hz)} & \textbf{Mass} & \textbf{Radius} \\
\hline
PSR J0952$-$0607 \cite{romani2022ApJ}  & Rotation-powered pulsar (RPP)  & 707 & $2.35 \pm 0.17$ M$_{\odot}$ & --- \\

PSR J0740+6620 \cite{riley2021}  & RPP & 347 & $2.072_{-0.066}^{+0.067}$ M$_{\odot}$ & $12.39_{-0.98}^{+1.30}$ km \\

PSR J0030+0451 \cite{riley2019}  & RPP  & 205 & $1.34_{-0.16}^{+0.15}$ M$_{\odot}$ & $12.71_{-1.19}^{+1.14}$ km \\ 

HESS J1731$-$347 \cite{doroshenko2022strangely}  & Central compact object & --- & $0.77_{-0.17}^{+0.20}$ M$_{\odot}$ & $10.4_{-0.78}^{+0.86}$ km \\

GW170817 (primary) \cite{abbott2019PhRvX} & BNS merger & --- & $1.46^{+0.12}_{-0.10}$ M$_{\odot}$ & $10.8^{+2.0}_{-1.7}$~km \\

GW170817 (secondary) \cite{abbott2019PhRvX} & BNS merger & --- & $1.27^{+0.09}_{-0.09}$ M$_{\odot}$ & $10.7^{+2.1}_{-1.5}$~km \\

 PSR J0737$-$3039A \cite{Kramer2006Sci} & Double pulsar (RPP) & 44 & $1.3381 \pm 0.0007$ M$_{\odot}$ & --- \\ 
\end{tabular}
\end{ruledtabular}
\label{tabConstraints}
\end{table*}

\Cref{tabConstraints} summarizes the observed pulsars used to constrain the free parameters of the DDQM and vector MIT bag models in \cite{daSilva:2023okq}. These include the massive, rapidly rotating pulsar PSR~J0952$-$0607~\cite{romani2022ApJ}, the NICER-measured millisecond pulsars PSR~J0740$+$6620~\cite{riley2021} and PSR~J0030$+$0451~\cite{riley2019}, and the low-mass compact object HESS~J1731$-$347~\cite{doroshenko2022strangely}. Four EOS parameterizations were derived for each model: (i) only the low-mass sources PSR~J0030$+$0451 and HESS~J1731$-$347 satisfy the constraints (DDQM1, MIT1); (ii) only the NICER pulsars PSR~J0740$+$6620 and PSR~J0030$+$0451 are satisfied (DDQM2, MIT2); (iii) all sources satisfy the constraints (DDQM3, MIT3); and (iv) only the high-mass pulsars PSR~J0952$-$0607 and PSR~J0740$+$6620 are satisfied (DDQM4, MIT4). In all cases, the maximum mass was constrained to \(2\,M_\odot \leq M_{\rm max} \leq 3.2\,M_\odot\) and causality was enforced. The parameter distributions consistent with the Bodmer--Witten conjecture for strange QM and the associated Bayesian posteriors are discussed in detail in \cite{daSilva:2023okq}. For DDQM, the priors were guided by Refs.~\cite{Wen:2005uf, Backes:2020fyw}, with \(C \in [-2,2]\) and \(\sqrt{D} \in [0,300]~\mathrm{MeV}\), while for the vector MIT bag model, following Refs.~\cite{Lopes_2021, lopes2022nature}, the explored ranges were \(B^{1/4} \in [130,170]~\mathrm{MeV}\), \(G_V \in [0,2]~\mathrm{fm}^2\), and \(b_4 \in [-2,2]\).

The parameter space considered in this work is constrained by a combination of observational and physical requirements. Observations of massive pulsars such as PSR~J0740+6620 ($M \approx 2.07\,M_\odot$) require that any viable EOS satisfies $M_{\max} \gtrsim 2\,M_\odot$. All selected parameterizations of both the vector MIT bag and DDQM models fulfill this condition (except DDQM1). Although not imposed explicitly, these parameter sets also support stable configurations at the canonical mass $1.4\,M_\odot$ (see \Cref{figmr}); the maximum mass of each model is, by construction, higher than this value. At the high-mass end, causality (sound speed $\leq c$ (constant speed of light)), together with the requirement of physical consistency at supranuclear densities, restricts the maximum mass to $\lesssim 3.2\,M_\odot$ for the parameter sets explored in the Bayesian analysis of \cite{daSilva:2023okq}. EOSs predicting significantly larger masses would require extreme stiffness approaching the causal limit and are therefore excluded. 

The Bodmer--Witten conjecture plays different roles in the two models. For the vector MIT bag model, it is imposed to ensure the absolute stability of strange quark matter at zero pressure, restricting the parameter space to combinations of $B$, $G_V$, and $b_4$ that yield an energy per baryon $\leq 930$ MeV. In contrast, our energy decomposition (see \Cref{figem}) shows that DDQM configurations are gravitationally bound ($U>0$), rather than self-bound. Consequently, in the DDQM case, the conjecture is not enforced as a binding condition but serves only as a qualitative guide. The dominant constraints instead arise from astrophysical observations (\Cref{tabConstraints}) and causality.

\begin{table*}[t]
 \centering
        \caption{Stellar properties of quark stars at their maximum gravitational mass ($M_{\max}$). The table lists the rotation parameter ($r_p/r_e$), the polar redshift ($Z_p$), the moment of inertia ($I$), the rotation frequency ($\nu$), the quadrupole moment ($Q$), the corresponding model parameters, and the percentage mass enhancement relative to the non-rotating case.}
    
    \begin{ruledtabular}
    \setlength\extrarowheight{2pt}
    \begin{tabular}{c c c c c c c c c c c c}
    \multicolumn{12}{c}{\textbf{DDQM model}} \\
    \hline
         Model & $C$ & $\sqrt{D}[{\rm MeV}]$ & & $r_p/r_e$ & M$_{\rm max}[{\rm M}_{\odot}]$ & $R[{\rm km}]$ & $I [10^{45}\rm g\,cm^2]$ & $Z_p$ & $\nu[{\rm Hz}]$ & $|Q|[10^{42}\rm g\,cm^2]$ & $\Delta M_{\rm max}$ [\%] \\
         \hline

                &  &  &  & 1.0 & 1.91 & 11.78 & 2.59 & 0.39 & 23 & 0.0 & 0.0 \\

            DDQM 1 & 0.50 & 137.5 & & 0.8 & 2.12 & 13.26 & 3.21 & 0.46 & 982 & 11.2 & 11.0 \\

                &  &  &  & 0.6 & 2.47 & 15.44 & 4.86 & 0.59 & 1299 & 17.8 & 29.3 \\

        \hline

                & & &  & 1.0 & 2.04 & 12.82 & 3.14 & 0.37 & 23 & 0.0 & 0.0 \\

            DDQM 2 & 0.65 & 132.2 &  & 0.8 & 2.27 & 14.27 & 3.91 & 0.46 & 912 & 12.2 & 11.3 \\

                &  &  &  & 0.6 & 2.64 & 16.92 & 6.21 & 0.56 & 1164 & 19.8 & 29.4 \\

        \hline

                &  &  &  & 1.0 & 2.10 & 13.25 & 3.33 & 0.37 & 19 & 0.0 & 0.0 \\

            DDQM 3 & 0.70 & 130.6 &  & 0.8 & 2.33 & 14.82 & 4.32 & 0.45 & 875 & 12.7 & 11.0 \\

                &  &  &  & 0.6 & 2.71 & 17.58 & 6.83 & 0.55 & 1117 & 20.7 & 29.1 \\

        \hline

                &  &  &  & 1.0 & 2.18 & 13.86 & 3.97 & 0.37 & 22 & 0.0 & 0.0 \\

            DDQM 4 & 0.80 & 127.4 & & 0.8 & 2.43 & 15.51 & 4.91 & 0.44 & 833 & 13.4 & 11.5 \\

                &  &  &  & 0.6 & 2.83 & 18.43 & 7.84 & 0.55 & 1060 & 22.0 & 29.8 \\

        \hline
        \hline
          \multicolumn{12}{c}{\textbf{vector MIT bag model}} \\
          \hline
         Model & $B^{1/4}[{\rm MeV}]$  &  $G_V[{\rm fm}^2]$  &  $b_4$ & $r_p/r_e$ & M$_{\rm max}[{\rm M}_{\odot}]$ & $R[{\rm km}]$ & $I [10^{45}\rm g\,cm^2]$ & $Z_p$ & $\nu[{\rm Hz}]$ & $|Q|[10^{42}\rm g\,cm^2]$ & $\Delta M_{\rm max}$ [\%] \\
        \hline

                &  &  &  &  1.0 & 2.21 & 11.66 & 3.31 & 0.50 & 27 & 0.0 & 0.0  \\

            MIT 1 & 140.90 & 0.116 & 0.72 & 0.8 & 2.45 & 12.85 & 3.91 & 0.63 & 1119 & 12.2 & 10.9 \\

                &  &  &  & 0.6 & 2.87 & 14.99 & 6.05 & 0.81 & 1452 & 19.8 & 29.9 \\
        
        \hline

                &  &  &  &  1.0 & 2.28 & 11.96 & 3.64 & 0.51 & 26 & 0.0 & 0.0   \\

            MIT 2 & 139.79 & 0.159 & 1.69 & 0.8 & 2.54 & 13.19 & 4.29 & 0.64 & 1096 & 12.7 & 11.4 \\

                &  &  &  & 0.6 & 2.97 & 15.35 & 6.58 & 0.83 & 1427 & 20.5 & 30.3 \\ 

        \hline

                &  &  &  &  1.0 & 2.40 & 12.46 & 4.16 & 0.51 & 23 & 0.0 & 0.0 \\

            MIT 3 & 137.96 & 0.235 & 1.63 & 0.8 & 2.67 & 13.75 & 4.92 & 0.65 & 1056 & 13.5 & 11.3 \\

                &  &  &  & 0.6 & 3.11 & 15.80 & 7.33 & 0.87 & 1409 & 21.4 & 29.6 \\
                
        \hline
         
                &  &  &  &  1.0 & 2.54 & 13.15 & 4.92 & 0.54 & 22 & 0.0 & 0.0  \\

            MIT 4 & 135.28 & 0.366 & 1.90 & 0.8 & 2.83 & 14.49 & 5.83 & 0.66 & 1005 & 14.4 & 11.4  \\

                &  &  &  & 0.6 & 3.30 & 16.65 & 8.68 & 0.88 & 1342 & 23.1 & 29.9 \\

    \end{tabular}
    \end{ruledtabular}
   
    \label{tabdd}
\end{table*}

\begin{table*}[t]
%\fontsize{6.6pt}{6.6pt}\selectfont
 \centering
        \caption{Stellar properties for the canonical configuration with mass ($1.4\,M_{\odot}$). Here, ($T/|W|$) denotes the ratio of rotational kinetic energy to gravitational binding energy.}
    
    \begin{ruledtabular}
    \setlength\extrarowheight{2pt}
    \begin{tabular}{ c c c c c c c c c c c }
    \multicolumn{11}{c}{\textbf{DDQM model}} \\
    \hline
         Model & $C$ & $\sqrt{D}[{\rm MeV}]$ & & $r_p/r_e$ & $R[{\rm km}]$ & $I [10^{45}\rm g\,cm^2]$ & $Z_p$ & $\nu[{\rm Hz}]$ & $|Q|[10^{42}\rm g\,cm^2]$ & $T/|W|$ \\
         \hline

                &  &  &  & 1.0 & 12.43 & 1.95 & 0.22 & 21 & 0.0 & $\cdots$ \\

            DDQM 1 & 0.50 & 137.5 & & 0.8 & 13.58 & 2.20 & 0.23 & 681 & 10.8 & 0.07 \\

                &  &  &  & 0.6 & 15.26 & 2.75 & 0.24 & 866 & 15.8 & 0.15 \\

        \hline

                & & &  & 1.0 & 13.37 & 2.23 & 0.20 & 20 & 0.0 & $\cdots$ \\

            DDQM 2 & 0.65 & 132.2 &  & 0.8 & 14.58 & 2.51 & 0.21 & 606 & 11.6 & 0.07  \\

                &  &  &  & 0.6 & 16.36 & 3.13 & 0.22 & 775 & 16.9 & 0.16 \\

        \hline

                &  &  &  & 1.0 & 13.86 & 2.35 & 0.19 & 17 & 0.0 & $\cdots$ \\

            DDQM 3 & 0.70 & 130.6 &  & 0.8 & 15.10 & 2.68 & 0.20 & 576 & 12.0 & 0.07 \\

                &  &  &  & 0.6 & 16.95 & 3.34 & 0.21 & 737 & 17.4 & 0.15 \\

        \hline

                &  &  &  & 1.0 & 14.38 & 2.47 & 0.18 & 20 & 0.0 & $\cdots$ \\

            DDQM 4 & 0.80 & 127.4 & & 0.8 & 15.66 & 2.87 & 0.19 & 540 & 12.4 & 0.07 \\

                &  &  &  & 0.6 & 17.54 & 3.57 & 0.20 & 694 & 18.1 & 0.16 \\

        \hline
        \hline
          \multicolumn{11}{c}{\textbf{vector MIT bag model}} \\
          \hline
         Model & $B^{1/4}[{\rm MeV}]$  &  $G_V[{\rm fm}^2]$  &  $b_4$ & $r_p/r_e$ & $R[{\rm km}]$ & $I [10^{45}\rm g\,cm^2]$ & $Z_p$ & $\nu[{\rm Hz}]$ & $|Q|[10^{42}\rm g\,cm^2]$ & $T/|W|$ \\
        \hline

                &  &  &  &  1.0 & 11.44 & 1.70 & 0.25 & 23 & 0.0 & $\cdots$  \\

            MIT 1 & 140.90 & 0.116 & 0.72 & 0.8 & 12.32 & 1.92 & 0.27 & 775 & 10.0 & 0.07 \\

                &  &  &  & 0.6 & 13.63 & 2.30 & 0.28 & 1013 & 14.3 & 0.15 \\
        
        \hline

                &  &  &  &  1.0 & 11.64 & 1.75 & 0.25 & 22 & 0.0 & $\cdots$ \\

            MIT 2 & 139.79 & 0.159 & 1.69 & 0.8 & 12.53 & 1.98 & 0.26 & 754 & 10.2 & 0.07 \\

                &  &  &  & 0.6 & 13.85 & 2.37 & 0.27 & 987 & 14.6 & 0.15 \\

        \hline

                &  &  &  &  1.0 & 11.94 & 1.83 & 0.24 & 21 & 0.0 & $\cdots$\\

            MIT 3 & 137.96 & 0.235 & 1.63 & 0.8 & 12.85 & 2.07 & 0.25 & 722 & 10.4 & 0.07 \\

                &  &  &  & 0.6 & 14.20 & 2.47 & 0.26 & 948 & 14.9 & 0.16 \\
                
    \hline
         
                &  &  &  &  1.0 & 12.36 & 1.90 & 0.23 & 20 & 0.0 & $\cdots$ \\

            MIT 4 & 135.28 & 0.366 & 1.90 & 0.8 & 13.31 & 2.20 & 0.24 & 681 & 10.8 & 0.07 \\

                &  &  &  & 0.6 & 14.69 & 2.63 & 0.25 & 899 & 15.4 & 0.16 \\

    \end{tabular}
    \end{ruledtabular}
   
    \label{tabdd1}
\end{table*}

\begin{figure}
    \centering
    \includegraphics[width=1\linewidth]{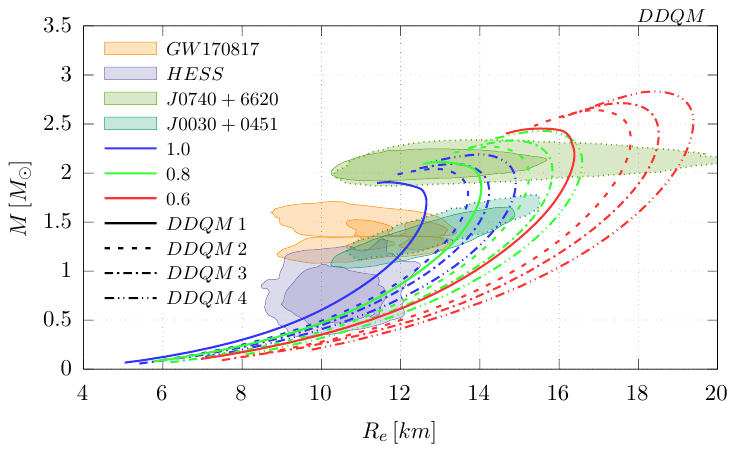}
    \includegraphics[width=1\linewidth]{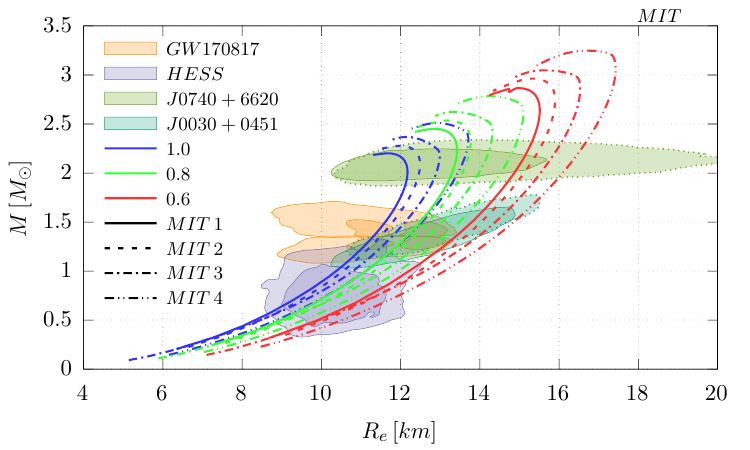}
    \caption{{Gravitational mass versus equatorial radius for the DDQM (top) and vector MIT bag (bottom) models. Different line styles correspond to different parameter sets, while colors indicate different ratios of the polar to equatorial radius. Confidence contours show observational constraints from the secondary component of the binary merger event GW170817 (orange; outer contour 90\% credible region (CR), inner contour 50\% CR), HESS~J1731--347 (light blue; outer contour 95\% CR, inner contour 68\% CR), PSR~J0740+6620 (green; dotted contour 95\% CR by Miller \textit{et al.}~\cite{miller2021}, solid contour 95\% CR by Riley \textit{et al.}~\cite{riley2021}, and PSR~J0030+0451 (blue; dotted contour 95\% credible region by Miller \textit{et al.}~\cite{Miller:2019cac}, solid contour 95\% CR by Riley \textit{et al.}~\cite{riley2019}).
}}
    \label{figmr}
\end{figure}

The structural properties of strange stars described by the DDQM and vector MIT bag models, for both static and rotating configurations, are summarized in Tables~\ref{tabdd} and~\ref{tabdd1}. \Cref{tabdd} corresponds to the maximum gravitational mass configurations, while \Cref{tabdd1} refers to sequences at the canonical mass of 1.4~$M_\odot$. The ratio of the polar-to-equatorial radius $r_p/r_e$ characterizes rotation, with $1.0$ denoting static stars and lower ratios associated with faster rotation. When comparing models calibrated against the same astrophysical constraints (i.e., DDQMi with MITi for the same index `i'), the stiffer vector MIT bag model yields systematically more massive and more compact stellar configurations than its DDQM counterpart. Static MIT configurations reach $M_{\max} \simeq 2.2$--2.5~$M_\odot$, compared to 1.9--2.2~$M_\odot$ for DDQM, while rapid rotation increases these limits to about 3.3 and 2.8~$M_\odot$, respectively. The DDQM stars exhibit larger equatorial radii (e.g., 18.4~km vs.\ 16.7~km for $r_p/r_e = 0.6$) and smaller moments of inertia and quadrupole moments, consistent with their lower compactness and gravitational binding. The MIT models, being stiffer, also sustain higher $\nu$ (up to $\sim 1450~\mathrm{Hz}$) than the DDQM models ($<1300~\mathrm{Hz}$).  

At the canonical mass of 1.4~$M_\odot$, some trends reverse. The DDQM EOS still produces stars with systematically larger radii (13.4--14.4~km) but, in this case, with moments of inertia (e.g., $I \approx 2.9\times10^{45}~\mathrm{g\,cm^2}$) and quadrupole moments are also higher than the more compact MIT configurations (11.4--12.4~km, $I \approx 2.2\times10^{45}~\mathrm{g\,cm^2}$). For both cases, maximum mass and canonical mass stars, the MIT model exhibits higher polar redshifts (e.g., $Z_p \sim 0.23$--0.25) relative to the DDQM models ($Z_p \sim 0.18$--0.22). The ratio of rotational kinetic to gravitational binding energy, $T/|W|$, remains comparable between the two models at a given $r_p/r_e$, implying similar rotational stability. Overall, the presence of a very massive ($M \gtrsim 2.5\,M_\odot$), rapidly rotating compact object would favor a stiffer EOS such as the vector MIT bag model, while canonical-mass pulsars with larger radii and moments of inertia would be more consistent with a softer EOS like the DDQM model. These complementary regimes jointly constrain the stiffness of QM and provide a discriminating test for competing EOSs.

The last column of \Cref{tabdd} shows that rotation robustly enhances $M_{\rm max}$ in all models. At moderate rotation ($r_p/r_e=0.8$), the increase is uniformly $\sim11\%$, while near the Kepler limit ($r_p/r_e=0.6$) it reaches $\sim29$–$30\%$ for both models, indicating that centrifugal support largely governs $\Delta M_{\rm max}$ rather than microphysical properties. In contrast, the frequencies required to achieve this enhancement differ: DDQM stars reach $M_{\rm max}$ at lower spins, whereas vector MIT stars require higher frequencies. Although the mass enhancement itself is nearly model-independent, the associated frequency range offers a potential observational discriminator between QM models. This enhancement exceeds the typical $\sim 20\%$ found for hadronic stars at Keplerian limit \cite{GourgoulhonA&A1999, ZhouAN2017, Gondek-RosinskaA&A2000}, consistent with universal relations predicting $\sim 15\!-\!20\%$ for hadronic matter \cite{Konstantinou:2022vkr} and $\sim 20\!-\!33\%$ for QSs \cite{Konstantinou_2026}. A direct comparison with \cite{daSilva:2025cfe}, which studied rotating stars composed of nucleonic, hyperonic, and $\Delta$-baryon matter, shows that purely nucleonic EOS yield a $\sim 9\!-\!13\%$ increase at the Keplerian limit, while the inclusion of $\Delta$-isobars leads to enhancements of up to $\sim 21\%$.

The gravitational mass as a function of the equatorial radius is shown in \Cref{figmr} for rotating strange stars modeled with the DDQM model (upper panel) and the vector MIT bag model (lower panel). In both cases, rotation shifts the sequences toward higher masses and larger radii due to centrifugal support. However, the magnitude of this shift depends strongly on the underlying QM EOS. The vector MIT bag model, being stiffer at high densities because of the repulsive vector interaction \cite{Lopes:2020btp, Lopes:2020dvs}, supports noticeably larger maximum masses near the Keplerian limit in the rotating regime. At the same time, the enhanced pressure leads to more compact configurations, yielding smaller radii for a given mass compared to the DDQM case. In contrast, the relatively softer DDQM EOS produces stars with larger radii and lower maximum masses under similar rotational conditions.

When comparing models with the same index and $r_p/r_e$ ratio near their respective Keplerian limits, rotation amplifies the intrinsic EOS differences: the stiffer vector MIT bag model can sustain more massive configurations, whereas the softer DDQM model reaches its Keplerian limit at relatively lower masses. Consequently, simultaneous measurements of mass, radius, and spin in rapidly rotating compact stars near the Keplerian limit, particularly from X-ray timing and future gravitational-wave observations, could provide a viable means of discriminating between these two QM scenarios. The observational confidence contours included in the plots highlight key compact objects: PSR J0740$+$6620 is shown in green \cite{riley2021, Miller:2021qha}, PSR J0030$+$0451 in blue \cite{riley2019, Miller:2019cac}, the secondary component of GW170817 in orange \cite{gw172018, LIGOScientific:2017vwq}, and the compact object in the supernova remnant HESS J1731$-$347 in purple \cite{Horvath:2023uwl}. The $M$ and $R_e$ of PSR~J0030+0451 and PSR~J0740+6620 are obtained from X-ray pulse profile modeling with NICER, while their spin frequencies ($\nu \simeq 205\,\mathrm{Hz}$ and $347\,\mathrm{Hz}$, respectively) are measured through radio timing. The observed sources shown in \Cref{figmr} are used as empirical benchmarks. They are not assumed to be QSs. Rather, their measured properties delineate regions of parameter space that any viable model must accommodate.

\begin{figure}
    \centering
    \includegraphics[width=1\linewidth]{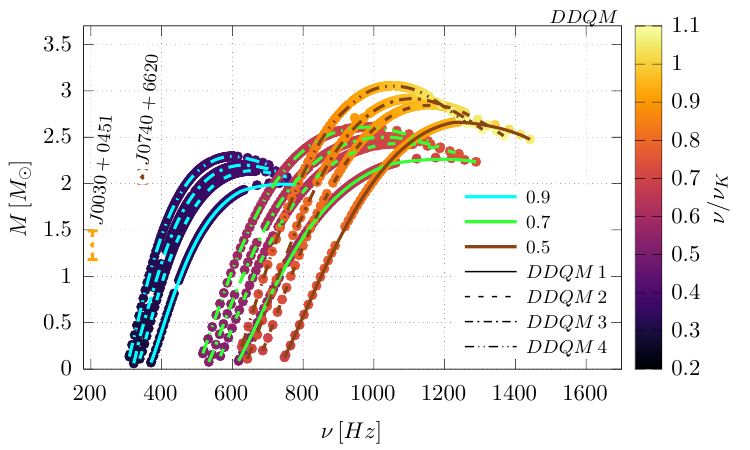}
    \includegraphics[width=1\linewidth]{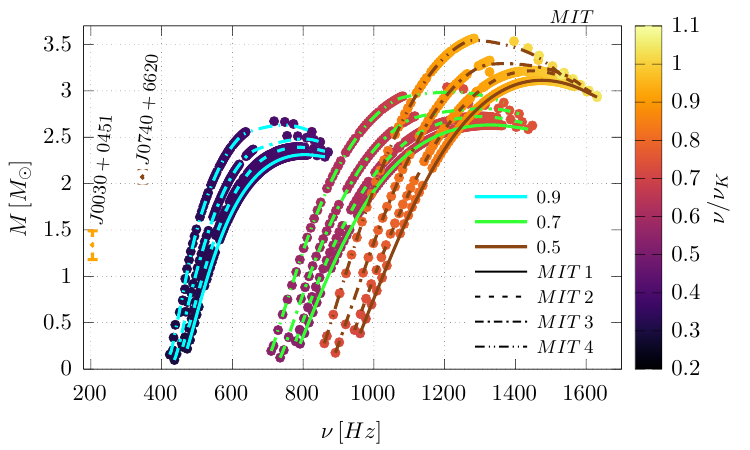}
    \caption{{Gravitational mass versus stellar frequency for the DDQM (top) and vector MIT bag (bottom) models. Different line styles correspond to different parameter sets, while line colors indicate different ratios of the polar to equatorial radius. The closed circles along the curves are color-coded according to the palette and represent the ratio of the spin frequency to the Keplerian frequency. Darker colors indicate lower rotation rates, and lighter colors indicate higher rotation rates, as shown in the color bar.}}
    \label{figmo}
\end{figure}

\Cref{figmo} shows the gravitational mass as a function of frequency for $\beta$ equilibrated strange star sequences constructed with the DDQM model (upper panel) and the vector MIT bag model (lower panel) for different ratios of $r_p/r_e$.  The color scale denotes the ratio $\nu / \nu_K$, which quantifies the proximity of each configuration to the Kepler frequency of a free particle in circular orbit at the equator. In both EOSs, rotation increases the maximum gravitational mass relative to the static limit, as centrifugal support partially counteracts gravitational collapse \cite{1967ApJ...150.1005H, Cook1994ApJ424}.  The spin and mass constraints of PSR J0740$+$6620 and PSR J0030$+$0451 indicate that both pulsars lie near the non-rotating limit ($r_p/r_e \approx 1.0$) or along slowly rotating sequences ($0.998 \lesssim r_p/r_e \lesssim 0.950$) for both EOS models. More strongly deformed configurations (e.g., $r_p/r_e = 0.9$) are excluded, as they would require spin frequencies well above the observed values.

The ratio $\nu/\nu_{K}$, indicated by the color bar in \Cref{figmo}, quantifies a star’s proximity to the Keplerian limit. Although it is not directly measurable, it can be inferred from joint constraints on $M$ and $R$, which together determine the Keplerian frequency $\nu_{K} \propto \sqrt{M/R^{3}}$. Thus, measurements of $M$ and $R$ from X-ray pulse profile modeling (e.g., with NICER \cite{riley2019, Miller:2019cac, riley2021, miller2021} or future missions such as enhanced X-ray Timing and Polarimetry (eXTP) \cite{watts2019dense, eXTP:2018anb}), combined with the observed spin frequency $\nu$ from radio timing, one can estimate the ratio $\nu/\nu_{K}$. An independent upper bound can be determined from the requirement, $\nu \leq \nu_{K}$, such that the maximum observed spin in a population constrains the typical values of $\nu/\nu_{K}$. 

Clear differences arise from the contrasting stiffness of the QM EOSs. The relatively stiffer vector MIT bag model (considering models with the same indices) supports significantly higher masses at all frequencies, reaching its maximum close to the Keplerian boundary. The dominance of moderate $\nu /\nu_K$ values indicates that stability is ultimately limited by general-relativistic collapse rather than equatorial Keplerian rotation \cite{Weber:2004kj}. The smaller radii of these compact configurations also enable higher Keplerian limits, allowing for stable rotation above $\sim$ 1500 Hz \cite{Paschalidis:2016vmz}. In contrast, the DDQM EOS yields a narrower stability domain, with lower maximum masses and rotational frequencies. Larger stellar radii reduce the Keplerian frequency, forcing DDQM stars to evolve close to the Keplerian threshold across most of the sequence. As reflected by the prevalence of high $\nu / \nu_K$ values, configurations with $\nu \gtrsim 1000$ Hz rapidly approach the Keplerian limit, restricting access to the extreme mass--frequency regime allowed by the MIT model, when models with the same index are compared.

These distinct behaviors have important observational implications. The detection of a very massive ($M \gtrsim 3\,M_\odot$) and rapidly rotating ($\nu \gtrsim 1400$ Hz) pulsar would strongly favor a stiff QM EOS such as the vector MIT bag model. Whereas the observed population of millisecond pulsars, with spin frequencies typically below $700\,\mathrm{Hz}$, is consistent with both models \cite{hessels2006Sci}. Consequently, the ratio $\nu/\nu_{K}$ becomes most diagnostic for the extreme high-frequency tail of the pulsar population.
 Current high-mass slowly rotating pulsars (e.g., PSR J0740$+$6620 \cite{NANOGrav:2019jur}) remain compatible with both scenarios, but future discoveries near the extremal mass--spin boundary will provide decisive constraints. Overall, the mass--frequency plane, supplemented by the rotational proximity indicator $\nu / \nu_K$, serves as a powerful probe of the stiffness of dense QM. The vector MIT bag model predicts a broader parameter space of stable rotating strange stars, while the DDQM model is restricted to lower masses and frequencies where Keplerian rotation dominates the stability limit.

\begin{figure}
    \centering
    \includegraphics[width=1\linewidth]{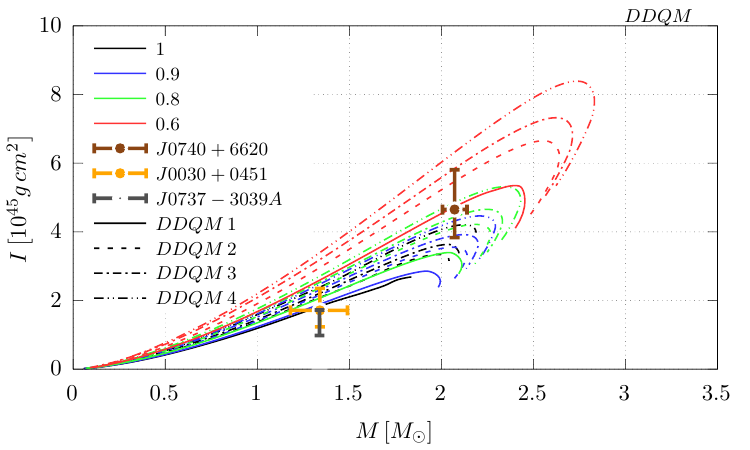}
    \includegraphics[width=1\linewidth]{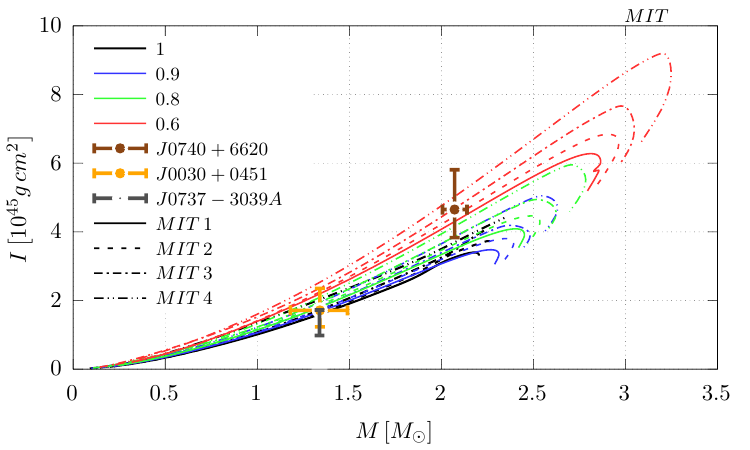}
    \caption{{Moment of inertia versus gravitational mass for the DDQM (top) and vector MIT bag (bottom) models. Different line styles correspond to different parameter sets, while colors indicate different ratios of the polar to equatorial radius. The observational constraints for PSR J0737$-$3039A \cite{burgay2003Natur, Bejger:2005jy}, PSR J0740$+$6620, and PSR J0030$+$0451 are indicated by data points with error bars.}}
    \label{figim}
\end{figure}

\Cref{figim} presents the moment of inertia ($I$) as a function of gravitational mass for rotating strange stars, computed with the DDQM EOS (top panel) and the vector MIT bag EOS (bottom panel). The moment of inertia, which quantifies the resistance to changes in rotational state, is a fundamental property of the star that depends critically on its internal mass distribution and, by extension, on the EOS \cite{lattimer2005ApJ,daSilva:2025cfe}. In both sequences, the \(I\) increases with $M$ because the combined growth of stellar mass and $R_e$ (for a given EOS) shifts more matter to larger distances from the rotation axis, thereby enhancing the $I$. The vector MIT bag model supports more massive, compact configurations, reaching its maximum $I$ at $M_{\rm max}$ before the onset of instability. In contrast, the DDQM EOS produces stars with larger radii at comparable masses, distributing mass farther from the rotational axis. This results in higher $I$ values at a fixed mass, as shown in \Cref{tabdd1}. Consequently, for a fixed mass, the DDQM model yields a higher \(I\), whereas at the maximum mass, the MIT model dominates due to its higher supported maximum mass.

It is important to mention that, for self-bound QM, the nearly constant internal energy density leads to the well-known non-rotating scalings $M \propto R^3$ and $I \propto MR^2\propto M^{5/3}$ \cite{1986ApJ...310..261A, Glendenning2000}. Rotation introduces quantitative deviations through centrifugal flattening and density redistribution \cite{1967ApJ...150.1005H}, but the near constant density throughout the stellar interior remains the primary origin of the strong correlations seen in \Cref{figmo} and \Cref{figim}. As shown in these figures, both the $\nu$ and $I$ increase with increasing mass up to the $M_{\rm max}$, before they start decreasing due to the onset of instability. 

Physically, the moment of inertia encodes the radial distribution of mass and pressure, providing a direct probe of QM stiffness and internal structure. Observationally, $I$ affects pulsar spin-down and orbital precession in binaries \cite{lattimer2005ApJ}. Additionally, precision pulsar timing in relativistic binaries, particularly those with massive components, offers prospects for direct $I$ determinations in the near future \cite{Stairs:2003eg,ozel2016ARA&A,Kramer:2021jcw}. Consequently, observing a large $I$ at high mass (see \Cref{tabdd}) would strongly favor stiff EOS like the vector MIT bag model, whereas a high $I$ around 1.4~$M_\odot$ would support a softer, more extended configuration consistent with DDQM predictions as shown in \Cref{tabdd1}. As multimessenger constraints improve, the $I$--$M$ relation emerges as a sensitive diagnostic for distinguishing competing QM models in rapidly rotating compact stars. In the figures, we show the estimated observable data from: orange for PSR J0030$+$0451~\cite{silva2021PhRvL}, brown for PSR J0740$+$6620~\cite{li2022CQGra}, and gray for PSR J0737$-$3039A~\cite{bejger2005MNRAS}.

\begin{figure*}
    \centering
    \includegraphics[width=0.45\linewidth]{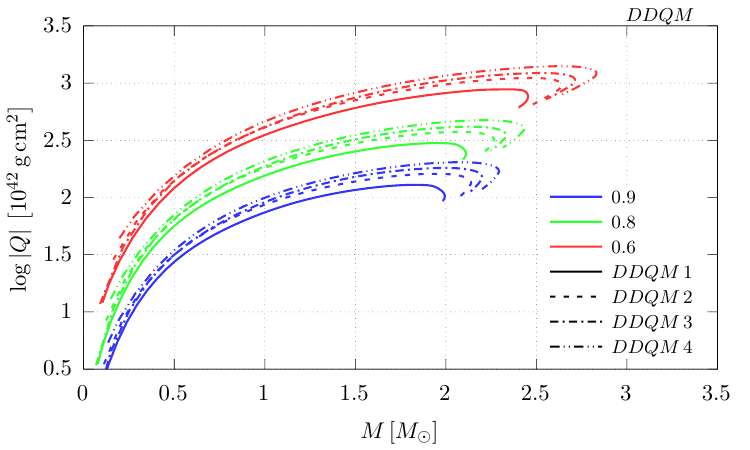}
    \includegraphics[width=0.45\linewidth]{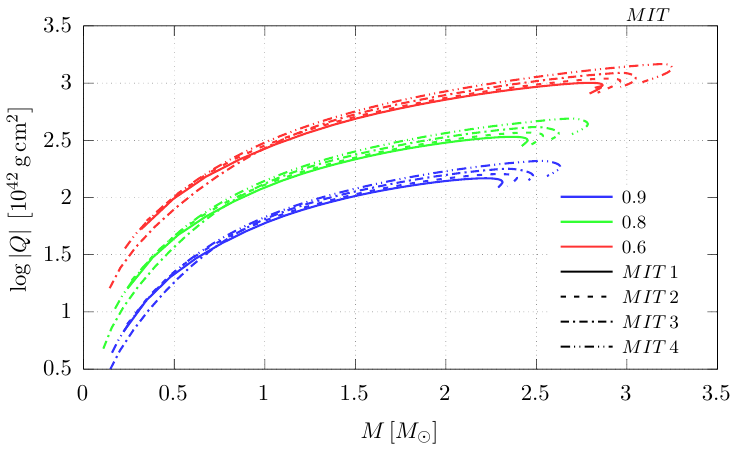}
    \caption{{Quadrupole moment versus gravitational mass for the DDQM (left) and vector MIT bag (right) models. Different line styles correspond to different parameter sets, while colors indicate different ratios of the polar to equatorial radius.}}
    \label{figqm}
\end{figure*}

The sensitivity of the quadrupole moment to the EOS is evident in \Cref{figqm}, where we plot it against the gravitational mass for rotating strange stars modeled with the DDQM EOS (left panel) and the vector MIT bag model (right panel). It quantifies the degree to which rotation deforms the star away from spherical symmetry and, consequently, the gravitational field outside the star, and directly encodes the stiffness or otherwise of the EOS. For stable stellar sequences (i.e., configurations below \(M_{\rm max}\) and $\nu \leq \nu_{K}$), the \(Q\) increases with mass for both EOSs; beyond \(M_{\rm max}\) and $\nu > \nu_{K}$, \(Q\) begins to decrease due to the onset of instability. Along sequences approaching the Kepler limit, more massive configurations also exhibit larger quadrupole moments, as their greater compactness allows them to sustain higher spins before reaching the Keplerian limit \cite{1967ApJ...150.1005H, Urbanec:2013fs, Paschalidis:2016vmz}. 

Despite this common trend, the models predict distinct behaviors. The vector MIT bag EOS produces more compact stars with stronger gravitational binding, which suppresses rotational flattening and results in systematically smaller quadrupole moments at a given mass \cite{Glendenning2000}. In contrast, the DDQM EOS leads to stars with larger radii (as shown in \Cref{figmr}) that are more easily deformed, yielding higher quadrupole moments \cite{Laarakkers:1997hb, Weber:2012ta} (see \Cref{tabdd1}). These differences highlight the strong sensitivity of the quadrupole moment to the internal stiffness of QM EOS. The quadrupole moment is a key parameter of the external spacetime, influencing phenomena such as the tidal response of compact objects in binaries, which affects the gravitational-wave signal during inspiral, and entering the I–Love–Q universal relations. Consequently, future multimessenger constraints that combine binary pulsar timing, X-ray observations, and gravitational-wave detections may provide a promising avenue for discriminating between these two QM scenarios \cite{Mashhoon:2006fj}. 

\begin{figure*}
    \centering
    \includegraphics[width=0.45\linewidth]{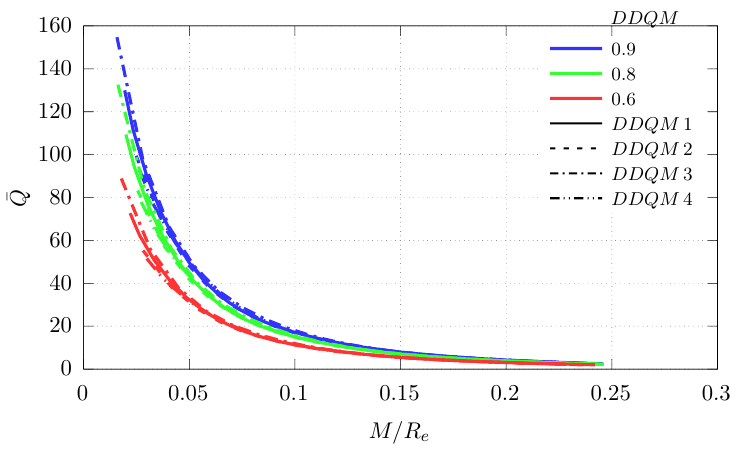}
    \includegraphics[width=0.45\linewidth]{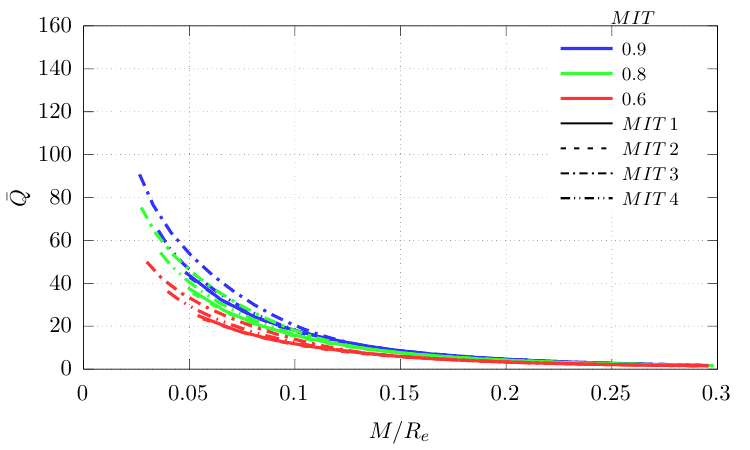}
    \caption{The dimensionless quadrupole moment versus compactness for the DDQM (left) and vector MIT bag (right) models. Different line styles correspond to different parameter sets, while colors indicate different ratios of the polar to equatorial radius.}
    \label{figqmC}
\end{figure*}

\Cref{figqmC} shows the dimensionless quadrupole moment $\bar{Q}=QM/J^{2}$ (geometric units) as a function of compactness $M/R_e$ (geometric units) for the DDQM (left) and vector MIT bag (right) models. As a scale-free measure of rotational deformation, $\bar{Q}$ enables direct comparison with the Kerr value ($\bar{Q}=1$) and removes explicit dependence on the absolute $M$ and $J$, allowing a consistent comparison across EOSs and rotation rates \cite{Urbanec:2013fs}. In all cases, $\bar{Q}$ decreases monotonically with increasing $M/R_e$, reflecting reduced deformability in more compact stars \cite{1967ApJ...150.1005H}. At fixed $M/R_e$ and $r_p/r_e$, DDQM configurations systematically yield larger $\bar{Q}$ than MIT ones, isolating the impact of EOS stiffness: the softer DDQM model is more deformable, while the stiffer MIT model produces smaller quadrupole responses. This behavior is consistent with the finding that the $\bar{Q}$–$M/R_e$ relation differs significantly between strange stars and NSs, providing a robust EOS discriminator \cite{Urbanec:2013fs}. Although parameter variations introduce some spread within each family, the separation between DDQM and MIT sequences remains clear over the full compactness range.

\begin{figure*}
    \centering
    \includegraphics[width=0.45\linewidth]{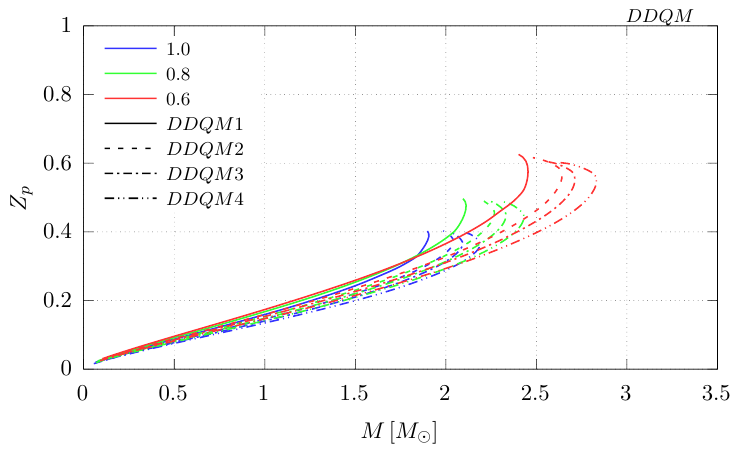}
    \includegraphics[width=0.45\linewidth]{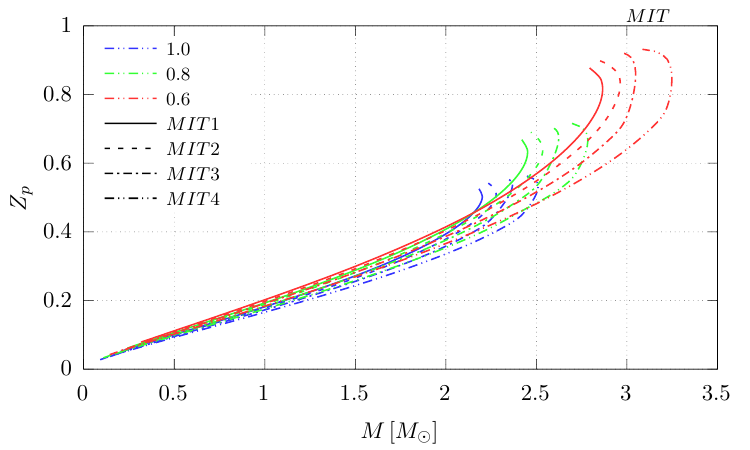}
    \caption{Polar redshift versus gravitational mass for the DDQM (left) and vector MIT bag (right) models. Different line styles correspond to different parameter sets, while colors indicate different ratios of the polar to equatorial radius.}
    \label{figzpm}
\end{figure*}

\Cref{figzpm} shows polar redshift versus gravitational mass for strange stars: the DDQM model (left panel) and the vector MIT bag model (right panel). Polar redshift directly reflects stellar compactness and is a key observable for strong-field gravity \cite{lattimer2001ApJ}. Both models show increasing redshift along the stable stellar mass sequences up to $M_{\rm max}$, but differ quantitatively. The vector MIT bag model exhibits a rapid rise, reaching redshifts $\gtrsim 0.85$ near maximum mass, indicating extremely compact configurations due to strong repulsive interactions \cite{Weber:2004kj}. The vector MIT bag model sequences show strong parameter dependence: MIT4 reaches $\gtrsim 0.85$, while MIT1–MIT3 peak between $\sim$ 0.6--0.7, indicating that only the most repulsive configurations achieve extreme compactness. In contrast, the DDQM model increases more gradually, reaching ~0.35 at $\sim 1.90\, M_\odot$ and ~0.55 at its maximum mass of $\sim 2.8\,  M_\odot$, which is systematically lower than all MIT sequences and reflects its larger radii and softer EOS behavior.

These differences have direct observational and theoretical implications. High redshifts from the MIT model would produce strong spectral line shifts detectable in X-ray observations \cite{Cottam:2002cu, Cottam:2007cd}, while the DDQM model predicts subtler effects. Redshift measurements above 0.7 would favor EOS-like MIT, whereas lower values are consistent with DDQM. The MIT sequences also illustrate parameter uncertainty, yet consistently predict high redshifts, highlighting strong pressure support against collapse. Overall, polar redshift serves as a sensitive probe of QM EOS stiffness. 

\begin{figure}
    \centering
    \includegraphics[width=1\linewidth]{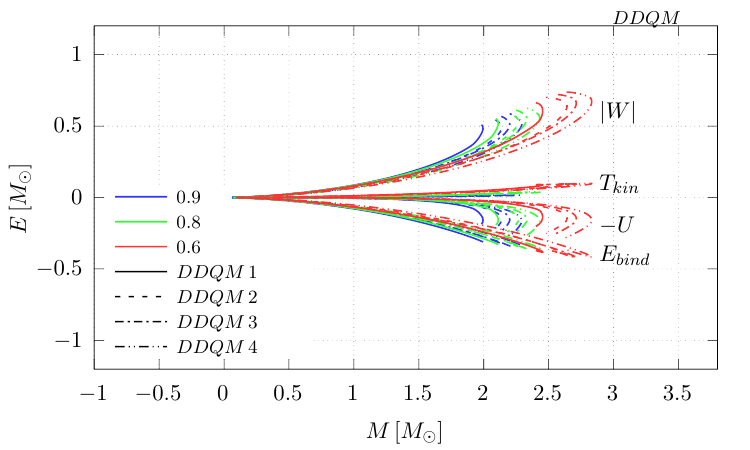}
    \includegraphics[width=1\linewidth]{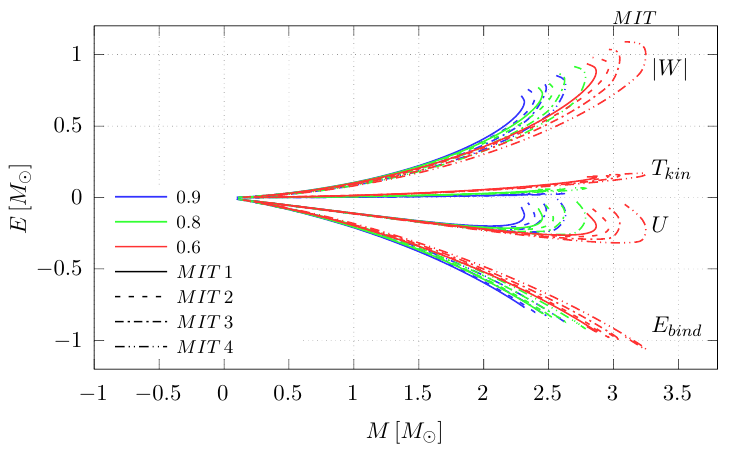}
    \caption{Four types of energy: gravitational energy $W$, kinetic energy $T$, internal energy $U$, and binding energy $E_{\mathrm{bind}}$, versus gravitational mass for the DDQM model on the upper panel, and for the vector MIT bag model on the bottom panel. The $U$ for the DDQM model is positive, in contrast to the vector MIT bag model, where $U$ is negative. Therefore, in the figures we plot $-U$ for the DDQM case to enable direct visual comparison between the two models.}
    \label{figem}
\end{figure}

\Cref{figem} presents the full energy decomposition for rotating strange stars, illustrating how $|W|$, $U$, $T$, and $E_\mathrm{bind}$ energies vary with gravitational mass. This illustrates how the EOS governs compactness and stability \cite{Weber:2004kj, daSilva:2025gdl}. The gravitational energy, $|W|$, dominates the total energy in both models, increasing steeply with mass and reflecting the strong self-gravity of these compact objects. The MIT sequences systematically exhibit larger $|W|$ values than their DDQM counterparts at comparable masses, directly highlighting the stiffer EOS, which produces more compact and gravitationally bound configurations. 

The internal energy $U$ exhibits a fundamental qualitative distinction between the two EOSs. In the vector MIT bag model, we observe a behavior that is not usual for NSs: $U$ remains negative and becomes more negative with increasing mass. This indicates a strongly self-bound strange QM state and can be seen as a direct consequence of Witten's conjecture for strange QM~\cite{Witten:1984rs}. In other words, such stars are ``energetically favorable'', much like atomic nuclei. The QSs are not merely gravitationally bound but are absolutely bound~\cite{Glendenning2000, Vartanyan1995Ap}. Conversely, in the DDQM model, $U$ is intrinsically positive (plotted with a minus sign only for comparison), reflecting a softer EOS in which density-dependent quark mass scaling reduces the binding contribution of the internal energy. This contrast demonstrates that the MIT model produces compact, tightly bound stars. The results show that $|W| \gg |U|$ in both models, which implies that the QM described by the MIT bag model ($U<0$) is intrinsically self-bound, whereas the resulting stellar configurations in both the MIT and DDQM models are bound predominantly by gravity, with DDQM stars being purely gravitationally bound.  

In contrast, the DDQM model yields more compressible, less strongly bound configurations  \cite{Weber:2004kj, daSilva:2023okq}, directly linking their microphysics to macroscopic stellar properties. In summary, the analysis of the star's energy budget reveals a critical difference: matter in the MIT model stars acts as though it's fundamentally `glued together' (self-bound), while matter in the DDQM model is held up more by its own pressure against gravity. This provides a clear, physical criterion to distinguish the two models. 

Rotational effects are crucial for precise pulsar timing measurements and play a significant role in structural deformation, NS evolution, and several observed phenomena~\cite{maeder2000ARA&A,ozel2016ARA&A,daSilva:2025cfe}. However, the rotational kinetic energy, $T$, remains smaller than $|W|$, emphasizing that the overall energy balance is primarily governed by gravity and strong interactions. We observed that the MIT models lead to higher values of $T$ than DDQM ones, since their EOSs allow the stars to reach higher rotational frequencies, as shown in \Cref{figmo}. 

As for the binding energy, both models exhibit negative $E_\mathrm{bind}$ across the entire mass range, and the absolute value of $E_\mathrm{bind}$ increases as the mass increases, which means the more massive the star, the more difficult it is to ``disassemble'' it. However, for a given mass, decreasing the value of the ratio $r_p/r_e$, and consequently, increasing the rotational frequency, leads to lower absolute values of  $E_\mathrm{bind}$, thus, rotation decreases the gravitational bound of the star. Comparing the two models, we can infer that the MIT models can lead to higher values of $E_\mathrm{bind}$, approximately twice as high as the DDQM models.

\subsection{Multimessenger signatures of QSs}
As shown in \Cref{tabdd1}, if a compact object with macroscopic properties (mass $\sim 1.34\,M_\odot$, spin $\sim 44$ Hz) similar to PSR J0737$-$3039A were a QS, a measurement of its $I$ with $\sim 10\%$ accuracy \cite{Bejger:2005jy} could help distinguish between the two QM models: values $I_{1.4} \gtrsim 2.16 \times 10^{45}\,\mathrm{g\,cm^2}$ would favor DDQM, whereas $I_{1.4} \lesssim 1.94 \times 10^{45}\,\mathrm{g\,cm^2}$ would support the vector MIT bag model. Additional constraints arise from gravitational-wave observations, which are indirectly sensitive to the stellar $Q$ through tidal interactions during the inspiral phase of binary systems. The DDQM stars are associated with larger $Q$ at a given $M$ and ratio $r_p/r_e$, compared to the stiffer vector MIT configurations. Complementarily, a detection of a large polar surface redshift, $Z_p \gtrsim 0.8$, in a massive and rapidly rotating pulsar would point toward an extremely compact object and would strongly favor the vector MIT model. We emphasize that the $Q$ discussed here is not directly observable, but can be inferred indirectly through tidal deformability measurements and universal I–Love–Q relations, rather than through continuous gravitational-wave emission. 

While both QSs and hadronic NSs exhibit similar qualitative trends, namely, the  $I$ and $Z_p$ increase with gravitational mass, their quantitative values differ significantly due to microphysical differences in their underlying EOS \cite{lattimer2001ApJ, Lattimer:2006xb}. Hadronic EOS, depending on their stiffness, produce more compact configurations with smaller $R$ at a given mass, leading to lower values of the $I$ \cite{lattimer2001ApJ}. In contrast, QM models, particularly softer ones such as DDQM, support stars with larger radii and consequently higher $I$ \cite{daSilva:2023okq}. Even more distinctive is the behavior of the $Z_p$, which directly probes stellar compactness. For example, pulsar profile modeling by \citet{Cottam:2002cu} inferred absorption-line redshifts of $Z_p = 0.35 \pm 0.05$ in NSs, while \citet{Miller:2019cac} estimated $Z_p \simeq 0.20$–$0.25$ for PSR~J0030+0451 and \citet{Miller:2021qha} found $Z_p \simeq 0.27$–$0.32$ for the more massive PSR~J0740+6620. In QSs, especially in stiff, self-bound EOS such as the vector MIT bag model, $Z_p$ can attain significantly larger values, with $Z_p \gtrsim 0.8$ becoming achievable. Such extreme redshifts are generally inaccessible to standard hadronic EOS \cite{Cottam:2002cu}, thereby providing a potentially clear observational signature for distinguishing QSs from hadronic NSs.

In particular, we find that the rotational enhancement of the $M_{\rm max}$ near the Kepler limit is systematically larger for QSs ($\sim 29\!-\!30\%$) than for typical hadronic stars ($\sim 20\%$ or less). The enhancement in the maximum mass of QSs occurs at characteristically higher spin frequencies than those supported by hadronic stars. In addition, the sign of the $U$, negative in the vector MIT model, reflecting intrinsically self-bound QM, and the large $Z_p$ attainable in stiff QM EOS, provide further microphysical and relativistic discriminants. When combined with simultaneous measurements of mass, radius, spin frequency, and moment of inertia from multimessenger observations, these features provide a coherent set of observational signatures. These signatures may help identify a compact object as a strange star, thereby offering a potential test of the strange matter hypothesis rather than merely enabling a comparison between different QM models. These observables are accessed through different astrophysical channels, as summarized below.

The $\nu$ is measured from radio timing of rotation-powered pulsars \cite{2004hpa..book.....L}, while the $M$ is obtained from relativistic binary timing (e.g., Shapiro delay) or X-ray pulse-profile modeling (NICER) \cite{riley2019, Miller:2019cac}. The $R$ is inferred primarily from X-ray observations \cite{ozel2016ARA&A}. Additional constraints arise from the $I$, accessible in double NS systems through relativistic spin--orbit coupling \cite{lattimer2005ApJ}, and from tidal deformability measured in gravitational-wave inspirals \cite{LIGOScientific:2017vwq, gw172018}. The $Q$ is not directly measurable for isolated pulsars but can be constrained indirectly through tidal effects and approximately universal $I$--Love--$Q$ relations \cite{Yagi:2013bca}. The surface redshift $Z_p$ may be inferred from spectroscopic observations \cite{Cottam:2002cu} or pulse-profile modeling.

Each observable probes complementary aspects of the stellar structure. Even when measured in different sources, they jointly restrict the EOS parameter space. In particular, $(M,R)$ measurements constrain the compactness, while an independent determination of $I$ breaks degeneracies between soft and stiff EOS. Mass--radius constraints set the scale of the $\nu_K$ \cite{HaenselA&A2009}, and measurements of the spin frequency $\nu$ further limit the maximum allowed ratio $\nu/\nu_K$. Gravitational-wave measurements add complementary constraints on tidal deformability and the quadrupole structure. Within this framework, the rotational observables computed here ($I$, $Q$, $Z_p$, $\Delta M_{\max}$, $\nu/\nu_K$) can be directly connected to measurable quantities, providing a consistent basis for testing QM EOS.

\section{Final Remarks and Conclusions} \label{remarks}

This work presents a comparative analysis of rotating compact stars using two representative QM EOS: the DDQM model and the vector MIT bag model. Our results (Tabs.~\ref{tabdd}--\ref{tabdd1}; Figs.~\ref{figmr}--\ref{figem}) show that rotation amplifies the intrinsic differences between these EOSs, providing a potentially clearer means of distinguishing their astrophysical imprints. For models calibrated against the same astrophysical data, the stiffer vector MIT model supports more massive and compact stars, reaching gravitational masses of up to \(\sim 3.3\,M_\odot\), compared with \(\sim 2.8\,M_\odot\) for the DDQM model, and can sustain higher Keplerian frequencies (\(\gtrsim 1450\,\mathrm{Hz}\)).

A combined study of $I$, $Q$, $Z_p$, $M$, $R_e$, $\nu$, and $E$ forms a strong basis for probing a dense matter. These observables are primarily governed by the stellar compactness and therefore do not exhibit a universal ordering between hadronic and QM EOSs. Depending on the stiffness and binding properties of the EOS, more compact configurations correspond to lower $I$ and $Q$, and higher $Z_p$. In contrast, QM models allow larger $R_e$ and higher $I$. In particular we found that soft EOS, such as the DDQM, produce relatively large $Q$, at intermediate masses ($\sim 0.5 - 1.5\, \rm M_\odot$), for models constrained by the same astrophysical data (i.e., models with identical indices), while stiff, QM EOS like the vector MIT bag model can reach extreme redshifts ($Z_p \gtrsim 0.8$). We emphasize that the two models studied here do not exhaust the possible QM EOS space, and current observations do not yet provide strong candidates for strange stars. Nevertheless, the distinct rotational signatures identified here, such as the larger enhancement of the $M_{\rm max}$ and the characteristic behavior of \(I\), \(Q\), \(\nu\), and \(Z_p\), define specific benchmarks that can be tested against future data from X-ray timing observations, binary pulsar campaigns, and gravitational-wave detectors.

The energy decomposition further underpins the microphysical distinction, revealing the self-bound character intrinsic to the MIT EOS (with negative $U$) in contrast with a completely gravitational bound DDQM (with positive $U$). Collectively, our results demonstrate that rotational data from multi-messenger astronomy will be important in establishing the potential existence and properties of strange QM. A coordinated analysis of the rotational properties ($I$, $Q$, $Z_p$, $M$, $R_e$, $\Delta M_{\rm max}$, and $\nu$) establishes a multi‑messenger framework for constraining the QM EOS. Such observations will help distinguish between the contrasting behaviors predicted by stiff, repulsion‑driven QM and softer, density‑dependent descriptions, as well as their hadronic matter counterparts, thereby informing the long-standing question of whether strange QM can be realized in nature.

Most of the analysis is based on comparisons between models constrained by the same astrophysical data. A brief, model-independent cross-comparison of the model parameters is therefore warranted. Independent of the specific parameter index, the DDQM and vector MIT bag models exhibit clear, systematic differences across their entire allowed ranges. For a fixed gravitational mass (e.g., $1.4\,M_\odot$), DDQM stars consistently have larger equatorial radii (by $\sim 1$--$2$ km) and higher $I$ (by $\sim 15$--$30\%$) than their vector MIT counterparts. The vector MIT family reaches significantly higher maximum masses ($\sim 3.3\,M_\odot$ vs. $\sim 2.8\,M_\odot$) and supports faster Keplerian rotation ($\sim 1450$ Hz vs. $\sim 1300$ Hz). Only vector MIT configurations can achieve large polar redshifts ($Z_p \gtrsim 0.8$) and exhibit negative internal energy ($U < 0$), indicating self-bound QM. In contrast, DDQM stars are purely gravitationally bound ($U > 0$), produce more moderate redshifts ($Z_p \lesssim 0.6$), and are more easily deformed, leading to larger $Q$ at a given spin. These trends persist across all parameter sets and are largely insensitive to the specific calibration, indicating that they arise from intrinsic differences in the underlying microphysics of the two models.

\begin{acknowledgments}

A.I. acknowledges financial support from the São Paulo State Research Foundation (FAPESP), Grant Nos. 2023/09545-1 and 2025/17347-0. 
This work is part of the project INCT-FNA (Proc. No. 464898/2014-5) and is also supported by the National Council for Scientific and Technological Development (CNPq) under Grants No.  306834/2022-7 (T.F.). T. F. also thanks the financial support from  Improvement of Higher Education Personnel CAPES (Finance Code 001) and FAPESP Thematic Grants (2023/13749-1 and 2024/17816-8). A.K. acknowledges financial support from “The three-dimensional structure of the nucleon from lattice
QCD” 3D-N-LQCD program, funded by the University of Cyprus, and from the projects HyperON (VISION ERC - PATH 2/0524/0001) and Baryon8 (POST-DOC/0524/0001), co-financed by the European Regional Development Fund and the Republic of Cyprus through the Research and Innovation Foundation.
F.M.S. would like to thank CNPq for financial support under research project No. 403007/2024-0 and research fellowship No. 201145/2025-1.
\end{acknowledgments}

\bibliographystyle{apsrev4-2}
\bibliography{reference1.bib}

\end{document}